\documentstyle[11pt,epsfig]{article}

\def\laq{~\raise 0.4ex\hbox{$<$}\kern -0.8em\lower 0.62
ex\hbox{$\sim$}~}
\def\gaq{~\raise 0.4ex\hbox{$>$}\kern -0.7em\lower 0.62
ex\hbox{$\sim$}~}

\begin{document}

\begin{titlepage}

\begin{flushright}
CERN-PH-TH/2007-229
\end{flushright}
\vspace*{1.8 cm}

\begin{center}

\huge
{Magnetogenesis, \\
spectator fields and CMB signatures}

\vspace{1cm}

\large{ Massimo Giovannini}\footnote{e-mail address: massimo.giovannini@cern.ch}\\
\normalsize
\vspace{.2in}
{\sl Centro ``Enrico Fermi", Compendio del Viminale, Via 
Panisperna 89/A, 00184 Rome, Italy}\\
\vspace{8mm}
{\sl Department of Physics, Theory Division, CERN, 1211 Geneva 23, Switzerland}

\begin{abstract}
A viable class of magnetogenesis models can 
be constructed by coupling the kinetic term of the hypercharge to 
a spectator field whose dynamics does not affect the inflationary 
evolution.  The magnetic power spectrum is explicitly related to the power spectrum of 
(adiabatic) curvature inhomogeneities when the quasi-de Sitter stage of expansion is driven by a single scalar degree of freedom. 
Depending upon the value of the slow-roll parameters,
the amplitude of smoothed magnetic fields over a (comoving) 
Mpc scale can be as large as $0.01$--$0.1$ nG at the epoch of the gravitational collapse of the 
protogalaxy.
 The contributions of the magnetic fields to the Sachs-Wolfe plateau and to the temperature autocorrelations in the Doppler region compare favourably 
 with the constraints imposed by galactic magnetogenesis.
Stimulating lessons are drawn on the interplay between magnetogenesis models and their possible CMB signatures.
\end{abstract}
\end{center}
\end{titlepage}
\newpage
Since the early fifties large-scale magnetic fields have 
inspired different areas of investigation  both at a theoretical and at a more phenomenological level
(see, as an example, Ref. \cite{zel1} for theoretical and historical 
accounts of the subject).
Both elliptical and spiral galaxies have magnetic fields at the $\mu$ G level \cite{gal1}. Abell clusters possess large-scale magnetic fields (not associated with individual galaxies) with typical correlation scale which can be as large as $100$ kpc \cite{clust1}. 
Superclusters have been also claimed 
to have magnetic fields \cite{supclust} at the $\mu$G level even if, in this case, crucial ambiguities 
persist on the way the magnetic field strengths are inferred from the Faraday rotation measurements.  The latest analyses of the AUGER experiment 
demonstrated a correlation between the arrival directions of cosmic rays 
with energy above $6\times10^{19}$ eV and the positions of active galactic 
nuclei within $75$ Mpc \cite{auger1}.  At smaller energies it has been 
convincingly demonstrated \cite{auger2} that overdensities on windows of $5$ deg radius (and for energies $10^{17.9} \mathrm{eV} < E < 10^{18.5} \mathrm{eV}$) are compatible 
with an isotropic distribution. Thus,  in the highest energy domain (i.e. energies larger  than $60$ EeV),
cosmic rays are not appreciably deflected: within a cocoon of $70$ Mpc the intensity of the (uniform) component of the putative magnetic field should be smaller than the nG.  On a theoretical ground, the existence of much larger  magnetic fields (i.e. ${\mathcal O}(\mu\mathrm{G})$) cannot be justified already  if
the correlation scale is of the order of $20$ Mpc.

In the late sixties Harrison \cite{harrison1} suggested that 
cosmology and astrophysics are just two complementary aspects of the origin of large-scale magnetic fields.   
Heeding observations there is no evidence against the primeval hypothesis 
even if the primordial origin of large-scale magnetism is not empirically compelling. 
Compressional amplification (taking place during the gravitational collapse of the protogalaxy) allows to connect the observed magnetic field to a protogalactic field, present prior to gravitational collapse, of typical strength ranging between  $0.1$  and $0.01$ nG.  
A better understanding of the interplay between dynamo  theory 
and the global conservation laws of magnetized plasmas
 has been recently achieved also because of the improved 
 comprehension of the solar dynamo action \cite{kanduaxel}. 
Within the dynamo hypothesis,  
the protogalactic field could be even much smaller than the nG and still explain some crucial properties of our magnetized Universe: astrophysical and cosmological mechanisms might really be complementary 
rather than mutually exclusive \cite{harrison1,kanduaxel} (see also \cite{zel1}, second reference).

The only direct way of putting the primordial hypothesis to test is represented by the observations related to the Cosmic Microwave Background \footnote{It is not excluded that 
the study of the morphological features of galactic fields will also give indications, albeit indirect, on the primordial nature of the protogalactic field (see Ref. \cite{gal1} and discussions therein).}
(CMB in what follows): the possible existence of a magnetized 
plasma prior to decoupling is germane to several CMB observables 
like temperature autocorrelations and cross-correlations (see \cite{max2} and references therein). Recently,
a semi-analytical technique has been developed
to compute more accurately than before the magnetized temperature and polarization autocorrelations 
as well as cross-correlations \cite{max3}: the gross logic of the method is 
to assume a dominant adiabatic mode in the pre-equality initial conditions and to  add, consistently, the effects of the magnetic fields in the Einstein-Boltzmann hierarchy and in the initial conditions. 

Large-scale magnetic fields produced 
inside the Hubble radius after inflation will have a correlation scale 
bounded (from above) by the Hubble radius at the moment 
when some charge separation is produced (be it, for instance, 
the electroweak time\footnote{This example holds under 
the assumption the electroweak phase transition is strongly first order 
which is, arguably, not the case.}). Since the Hubble radius, during radiation, evolves much faster than the correlation scale of the produced field, the typical scale over which the magnetic field is coherent today is 
much shorter than the Mpc, obliterating, in this way, the 
possibility of successfully reproducing the galactic magnetic field 
\cite{max2}. Conversely, the physical rationale for 
inflationary magnetogenesis resides on the possibility of achieving 
large correlation scales:  quantum fluctuations of Abelian gauge 
fields can be amplified in the same way as  zero point fluctuations of the geometry are amplified.  Unlike the scalar and tensor modes of the geometry,  Abelian gauge fields (like the hypercharge) in four space-time dimensions obey Weyl invariant evolution equations \cite{weyl1}.
Since the pumping action of the background geometry is not efficient in amplifying the fluctuations of gauge fields,
 Weyl invariance should  be broken for the viability of the whole 
 construction \cite{weyl1}.   
 The amplified gauge fields should be Abelian. The only non-screened 
vector modes that are present at finite conductivity 
are the ones associated with the hypercharge field \cite{max4}.
The non-Abelian fields develop actually a mass and they are screened as the Universe thermalizes.
After the electroweak phase transition the photon field remains unscreened 
with amplitude $\cos{\theta_{\mathrm{w}}} \vec{{\mathcal Y}}$.
While the coupling of the hypercharge to fermions is chiral, the QED coupling is vector-like. At 
finite conductivity, however, the descriptions of the two plasmas are similar
\footnote{See \cite{max4} and the equations of anomalous magnetohydrodynamics, i.e. the generalization of magnetohydrodynamics to the case where anomalous effects are included.} 
and can be given in terms of an effective (Ohmic) current which is proportional 
to the (hyper)electric field. The specific nature of the gauge field is often ignored in the current literature:  the main endevour is to break consistently Weyl invariance (possibly maintaining gauge invariance).  The  
Abelian field arising in this case which should be thought, indeed, as a putative hypercharge field.

In the present paper it will be argued that Weyl invariance can be broken 
through the  coupling of a spectator field to the gauge kinetic term also in the case of conventional inflationary scenarios.
A spectator field is defined, in the present context, as a field which does not drive the inflationary 
evolution but which is, nonetheless, dynamical. It is not excluded, in the present context,
 that the resulting large-scale magnetic fields are amplified to nG strength and with a
nearly scale-invariant spectrum. 
 The field content of the model is apparent from the total action which 
 includes, on top of the gravitational part, the contribution 
 of the inflaton $\varphi$ and of the spectator field $\psi$:
\begin{equation}
S_{\mathrm{tot}} = S_{\mathrm{gravity}} + S_{\varphi} + S_{\psi}.
\label{totac}
\end{equation}
The various components of the total action can be written, in explicit terms, as\footnote{The conventions on the four-dimensional  metric will be mostly minus, i.e. $(+,-,-,-)$. Recall also that 
$\overline{M}_{\mathrm{P}} = M_{\mathrm{P}}/\sqrt{8\pi}$ with $M_{\mathrm{P}} = 1.22\times 
10^{19} \mathrm{GeV}$.}
\begin{eqnarray}
&&S_{\mathrm{gravity}} =-\frac{\overline{M}_{\mathrm{P}}^2}{2} \int d^{4} x \sqrt{-g} R,
\qquad S_{\varphi} = \int d^{4} x \sqrt{-g} \biggl[\frac{1}{2} g^{\alpha\beta}\partial_{\alpha}\varphi \partial_{\beta} \varphi - V(\varphi)\biggr],
\label{grphiac}\\
&& S_{\psi} = \int d^{4} x \sqrt{-g} \biggl[\frac{1}{2} g^{\alpha\beta}\partial_{\alpha}\psi \partial_{\beta} \psi - W(\psi) - \frac{\lambda(\psi)}{16 \pi} Y_{\alpha\beta}
Y^{\alpha\beta} \biggr],
\label{psiac}
\end{eqnarray}
where $V(\varphi)$ and $W(\psi)$ are, respectively, the inflaton potential 
and the potential of the spectator field. The hypercharge field strength$Y_{\alpha\beta} = \nabla_{[\alpha} Y_{\beta]}$ 
is defined in terms of the covariant derivative with respect to the four-dimensional metric $g_{\mu\nu}$. 
In Eq. (\ref{psiac}), $\lambda(\psi)$ denotes 
the coupling of $\psi$ to the hypercharge field. 

In a conformally flat Friedmann-Robertson-Walker  metric $g_{\mu\nu} = a^2(\tau) \eta_{\mu\nu}$ (where 
$\eta_{\mu\nu}$ is the four-dimensional Minkowski metric), 
the Hamiltonian constraint stemming from the equations derived from the total action 
(\ref{totac}) is given by
\begin{equation}
\overline{M}_{\mathrm{P}}^2 {\mathcal H}^2 = \frac{1}{3}\biggl[\frac{{\varphi'}^2}{2} + V a^2\biggr] 
+ \frac{1}{3}\biggl[\frac{{\psi'}^2}{2} + W a^2\biggr]  + \frac{1}{8\pi} (\vec{{\mathcal B}}^2 + 
\vec{{\mathcal E}}^2),
\label{HAM1}
\end{equation}
where the prime denotes the derivation with respect to the conformal time coordinate $\tau$ and 
${\mathcal H} = a'/a$ is related to the Hubble parameter $H$ as ${\mathcal H} = H/a$. In Eq. (\ref{HAM1}) $\vec{{\mathcal E}} = \sqrt{\lambda} \,\vec{e}$ and $
\vec{{\mathcal B}} = \sqrt{\lambda} \,\vec{b}$ are, respectively, the 
hyperelectric and the hypermagnetic fields defined, from the field strength as\footnote{The 
rescaling of $\vec{e}$ and $\vec{b}$ through $\sqrt{\lambda}$ arises since 
the hypercharge energy-momentum tensor contains the coupling to $\psi$ throught $\lambda$. These will not be, however, the normal modes of the system as it will be clear in a moment.}
$Y_{0i} = a^2 e_{i}$ and $Y_{ij} = - a^2 \epsilon_{ijk} b^{k}$.
The dual field strengths (appearing in the Bianchi identity) are simply $\tilde{Y}_{ij} = a^2 e^{m} \epsilon_{mij}$ and $\tilde{Y}_{0i} = a^2 b_{i}$.
 The field $\varphi$ is the dominant source of the background geometry while 
 $\psi$ is a spectator field which is allowed to roll during inflation but which gives a negligible 
 contribution to the background geometry. Denoting by $\psi_{i}$ the initial value of $\psi$ 
 at a curvature scale $H_{i}$ this requirement implies that 
 \begin{equation}
\psi_{\mathrm{i}}^2 <  \frac{2}{3} \biggl(\frac{H_{1}}{H_{\mathrm{i}}}\biggr)^2 \overline{M}_{\mathrm{P}}^2
\label{cond1}
\end{equation}
where $H_{1}$ is the curvature scale at the end of inflation. When $\psi$ starts rolling at 
$\tau_{i}$ the hyperelectric and the hypermagnetic fields are just given by 
their corresponding quantum fluctuations and are therefore even smaller than the 
energy density of $\psi$.  The (homogeneous) 
evolution equation for $\psi$ will therefore be given by 
\begin{equation}
 \psi'' + 2  {\mathcal H} \psi'  +  \frac{\partial W}{\partial\psi} a^2 + \frac{a^2}{8\pi} \frac{\partial \ln{\lambda}}{\partial \psi}(\vec{{\mathcal B}}^2 + \vec{{\mathcal E}}^2)=0.
\label{psieq}
\end{equation}
The values of the hypermagnetic (and hyperelectric) fields 
generated from quantum fluctuations will be always smaller than the energy density of $\psi$. This implies that the back-reaction terms arising, for instance, in Eqs. (\ref{HAM1}) and (\ref{psieq}) can be safely neglected. 
It will be assumed that $W(\psi) = m^2 (\psi -\psi_{*})/2$ with $m < H_{1}$. 
Since $\psi$ is light during inflation, it will also be required that
$\psi_{*} \ll M_{\mathrm{P}}$. 
Deep in the course of the inflationary epoch the evolution 
equation of $\psi$ is then
\begin{equation}
\Sigma'' +[ \mu^2 - (2 - \epsilon)] a^2 H^2 \Sigma =0,\qquad 
\Sigma = a \psi,\qquad \epsilon= - \frac{\dot{H}}{H^2} = \frac{\overline{M}_{\mathrm{P}}}{2} \biggl(\frac{V_{,\varphi}}{V}\biggr)^2
\label{PSI2}
\end{equation}
having introduced $\mu=m/H$ and the first slow-roll parameter $\epsilon$ 
which is related to the first derivative of the inflaton potential.
In the limit $\mu\ll 1$ the evolution of $\psi$ will be simply given by
\begin{equation}
\psi = \psi_{\mathrm{i}}\biggl( - \frac{\tau}{\tau_{\mathrm{i}}}\biggr)^{\beta} + \psi_{*},\qquad \beta = \frac{3 - 2 \epsilon}{1 - \epsilon}.
\label{PSI3}
\end{equation}
As the field $\psi$ evolves in time, the hypermagnetic and hyperelectric 
fields can be parametrically amplified, as it follows from the 
equations of motion easily obtainable by the appropriate functional variation of the 
total action (\ref{totac}):
\begin{equation}
\nabla_{\mu} (\lambda Y^{\mu\nu}) = 4\pi J^{\nu}, \qquad \nabla_{\mu} (\tilde{Y}^{\mu\nu})=0,\qquad 
\lambda(\psi) = \biggl(\frac{\psi - \psi_{*}}{\overline{M}_{\mathrm{P}}}\biggr)^{\alpha}.
\label{Y1}
\end{equation}
where the contribution of the (Ohmic) current $J^{\nu}$ has been 
included for convenience. In Eq. (\ref{Y1}) 
the expression of $\lambda(\psi)$ contains the parameter 
$\alpha$ which will eventually determine the slope of the gauge field 
spectra and which will be constrained by phenomenological considerations.
In the conformally flat metric $g_{\mu\nu} = a^2(\tau) \eta_{\mu\nu}$, Eq. (\ref{Y1}) can be written, using vector notations, as:
\begin{eqnarray}
&&\vec{\nabla} \times ( a^2 \sqrt{\lambda} \vec{{\mathcal B}}) = \frac{\partial}{\partial\tau} [ a^2 \sqrt{\lambda} 
\vec{{\mathcal E}}] + 4\pi \vec{J},\qquad \vec{\nabla}\cdot\vec{J} =0,
\label{Y2}\\
&& \frac{\partial}{\partial \tau} \biggl[ \frac{a^2 \vec{\mathcal B}}{\sqrt{\lambda}}\biggr] + 
\vec{\nabla}\times \biggl[ \frac{a^2 \vec{\mathcal E}}{\sqrt{\lambda}}\biggr]  =0.
\label{Y3}
\end{eqnarray}
where $\vec{J} = a^3 \sigma_{\mathrm{c}} = \sigma a^2 \vec{e} = \sigma a^2 \vec{{\mathcal E}}/{\sqrt{\lambda}}$;
$\sigma(\tau) = \sigma_{\mathrm{c}} a(\tau)$ denotes the rescaled value of the conductivity 
and it appears because of the choice of the conformal time coordinate as a pivot variable of the 
system.  Since $\lambda$ depends only upon $\tau$ the Ohmic current is always 
divergence-less as it should be by definition.  
Combining Eqs. (\ref{Y2}) and (\ref{Y3}) in the absence 
of conductivity (i.e. during inflation) the hypermagnetic and hyperelectric fields obey the following pair 
of (decoupled) equations:
\begin{eqnarray}
 \vec{B}'' - \nabla^2 \vec{B} - \frac{(\sqrt{\lambda})''}{\sqrt{\lambda}} \vec{B} =0,\qquad \vec{E}'' - \nabla^2 \vec{E} - \sqrt{\lambda} \biggl(\frac{1}{\sqrt{\lambda}}\biggr)'' \vec{E} =0,
\label{hmagn}
\end{eqnarray}
where $\vec{E} = a^2 \vec{{\mathcal E}}$ and $\vec{B} = a^2 \vec{{\mathcal B}}$ are the normal modes of the system. The dual nature of the pump fields for $\vec{E}$ and $\vec{B}$ in Eq. (\ref{hmagn}) is a reflection of the strong-weak coupling duality 
of the Abelian theory in the absence of sources (see, for instance, \cite{sdual}).
During inflation the gauge field fluctuations can then be quantized in the Coulomb 
gauge (which is the appropriate one for 
treating gauge fields in time-dependent background geometries \cite{lford}) and the vector potential 
can be expanded in terms of the appropriate mode functions $f_{k}(\tau)$
\begin{equation}
 \hat{{\mathcal Y}}_{i}(\vec{x},\tau) = \frac{1}{(2\pi)^{3/2}} \sum_{\gamma} e^{(\gamma)}_{i}
\int d^{3} k [ \hat{a}_{\vec{k},\gamma} f_{k}(\tau) e^{- i \vec{k}\cdot \vec{x}} +\hat{a}_{\vec{k},\gamma}^{\dagger} f_{k}^{*}(\tau) 
e^{ i \vec{k}\cdot \vec{x}} ],
\label{vecpot}
\end{equation}
where $e^{(\gamma)}_{i}$ is the polarization unit vector; $\hat{a}_{k,\gamma}$ and $\hat{a}_{\vec{k},\gamma}^{\dagger}$ obey $[\hat{a}_{k,\gamma}, \hat{a}_{\vec{p},\gamma}^{\dagger}] = \delta_{\gamma\gamma'} \delta^{(3)}(\vec{k} - \vec{p})$.
Since $ \vec{B} = \vec{\nabla}\times \vec{{\mathcal Y}}$, the mode function $f_{k}(\tau)$ (and its complex conjugate) will satisfy the same equation obeyed by $\vec{B}$ (see Eq. (\ref{hmagn})). 

At end of inflation the Universe reheats. Thanks to the decay 
of the inflaton and of the spectator field the quasi-de Sitter 
background becomes effectively dominated by a fluid of ultra-relativistic 
particles with radiative equation of state. Overall the plasma is
 globally neutral but the conductivity becomes 
 large since charged species are copiously produced.
Lorentz invariance is then  broken and hyperelectric fields are strongly
suppressed while the hypermagnetic fields survive. 
A preferred physical frame naturally emerges, 
i.e. the so-called plasma frame 
where the conductivity is finite and the hyperelectric fields are dissipated. 
Since $\psi$ decays, $\lambda$ will freeze and 
the system of Eqs. (\ref{Y2}) and (\ref{Y3}) can be written as
\begin{equation}
\frac{\partial \vec{E}_{\mathrm{a}}}{\partial \tau} +4 \pi \sigma \vec{E}_{\mathrm{a}}
= \vec{\nabla}\times \vec{B}_{\mathrm{a}},\qquad \frac{\partial \vec{B}_{\mathrm{a}}}{\partial \tau} = - \vec{\nabla}\times \vec{E}_{\mathrm{a}},
\label{M2}
\end{equation}
where the subscript ``a" signifies that the hyperelectric and hypermagnetic fields are computed after the 
transition to radiation.
 Denoting with the subscript ``b" the field variables after the rise of the conductivity  the 
 appropriate continuity conditions for the magnetic and the electric fields are:
\begin{equation}
\vec{B}_{\mathrm{a}} = \vec{B}_{\mathrm{b}},\qquad 
\vec{E}_{\mathrm{a}} = \frac{\vec{\nabla}\times \vec{B}_{\mathrm{a}}}{4\pi \sigma} = 
 \frac{\vec{\nabla}\times \vec{B}_{\mathrm{b}}}{4\pi \sigma}.
\label{M3}
\end{equation}
 Equation (\ref{M3}) stipulates that, after the transition, the electric fields 
are suppressed by the conductivity as soon as radiation dominates.
Solving Eq. (\ref{hmagn}) during inflation and Eq. (\ref{M2}) during radiation
the boundary conditions (\ref{M3}) permit the estimate of the two point function of the hypermagnetic field operators for a generic time $\tau > \tau_{1}$ where $\tau_{1}$ denotes the epoch of the sudden rise in the conductivity:
\begin{equation}
\langle 0| \hat{B}_{i}(\vec{x},\tau) \hat{B}_{j}(\vec{y},\tau) |0 \rangle 
= \int d \ln{k} \,\, P_{B}(k) P_{ij}(k) \frac{\sin{kr}}{kr},\qquad r =|\vec{x} - \vec{y}|,
\label{M15}
\end{equation}
where $P_{\mathrm{B}}(k)$ and $P_{ij}(k)$  denote, respectively, the hypermagnetic power spectrum and the traceless projector
\begin{equation}
P_{B}(k) = {\mathcal C}(\delta) H_{1}^4  \biggl(\frac{k}{k_{1}}\biggr)^{n_{\mathrm{B}-1}} e^{- 2 \frac{k^2}{k_{\sigma}^2}},\qquad P_{ij}(k)= \biggl( \delta_{ij} - \frac{k_{i} k_{j}}{k^2}\biggr).
\label{M16}
\end{equation}
In Eq. (\ref{M16}) ${\mathcal C}(\delta) = 2^{2\delta-1} \Gamma^2(\delta)/\pi^2$ and 
\begin{equation}
\delta =  \frac{3\alpha -1 + \epsilon( 1 - 2 \alpha)}{2 ( 1 - \epsilon)},\qquad n_{\mathrm{B}} 
= \frac{7 - 3\alpha - \epsilon ( 7 - 2\alpha)}{1 - \epsilon},\qquad k_{1} = \frac{1}{\tau_{1}}.
\label{M16a}
\end{equation}
In Eq. (\ref{M16}) $k_{\sigma}$ is the conductivity wave-number, i.e. 
$k_{\sigma}^{-2} = \int d\tau/(4\pi\sigma)$.
The wave-numbers $k_{1}$ and $k_{\sigma}$ can be also usefully expressed, within a comoving coordinate system, in 
units of $\mathrm{Mpc}^{-1}$:
\begin{equation}
k_{1} = 1.1\times 10^{24} (\epsilon {\mathcal P}_{\mathcal R})^{1/4}\,\,\mathrm{Mpc}^{-1},\qquad 
{\it k}_{\sigma} = 1.55 \times 10^{12} 
\biggl( \frac{ h_{0}^2 \Omega_{\mathrm{b}0}}{0.023}\biggr)^{1/2} \biggl(\frac{h_{0}}{0.7}\biggr)^{1/2} \, \mathrm{Mpc}^{-1},
\label{scales}
\end{equation}
where $\epsilon$ is the slow-roll parameter already encountered in Eq. (\ref{PSI3}) and ${\mathcal P}_{\mathcal R}
\simeq 2.35\times 10^{-9}$ is the inflationary power spectrum of curvature perturbations evaluated at 
the pivot scale $k_{\mathrm{p}} = 0.002\,\mathrm{Mpc}^{-1}$ and estimated according to the WMAP data 
alone \cite{WMAP}.
The wave-numbers of Eq. (\ref{scales}) indeed correspond to very short wavelengths as it can be appreciated 
by comparing them to the comoving wave-number corresponding to the Hubble radius, i.e. 
$k_{0} = 2.33 \times 10^{-4} (h_{0}/0.7)\,\, \mathrm{Mpc}^{-1}$. The spectrum of Eq. (\ref{M16}) holds 
for $k < k_{1}$. But since the exponential fall-off triggered by the finite value of the conductivity 
becomes relevant already at $k\simeq k_{\sigma}$ the power-law 
behaviour is only verified for sufficiently small wave-numbers $k < k_{\sigma}$.
The two-point function of Eq. (\ref{M15}) has been computed by 
quantizing the system in the Coulomb gauge and by solving the resulting evolution equations in the Heisenberg representation. The final 
result (\ref{M15}) can also be expressed as a statistical condition on the (classical) Fourier amplitudes 
\begin{equation}
\langle B_{i} (\vec{k}) B_{j}^{*}(\vec{p})\rangle = \frac{2\pi^2}{k^3} P_{B}(k) P_{ij}(k)
\delta^{(3)}(\vec{k} - \vec{p}).
\label{M17}
\end{equation}
\begin{figure}
\begin{center}
\begin{tabular}{|c|c|}
      \hline
      \hbox{\epsfxsize = 6.9 cm  \epsffile{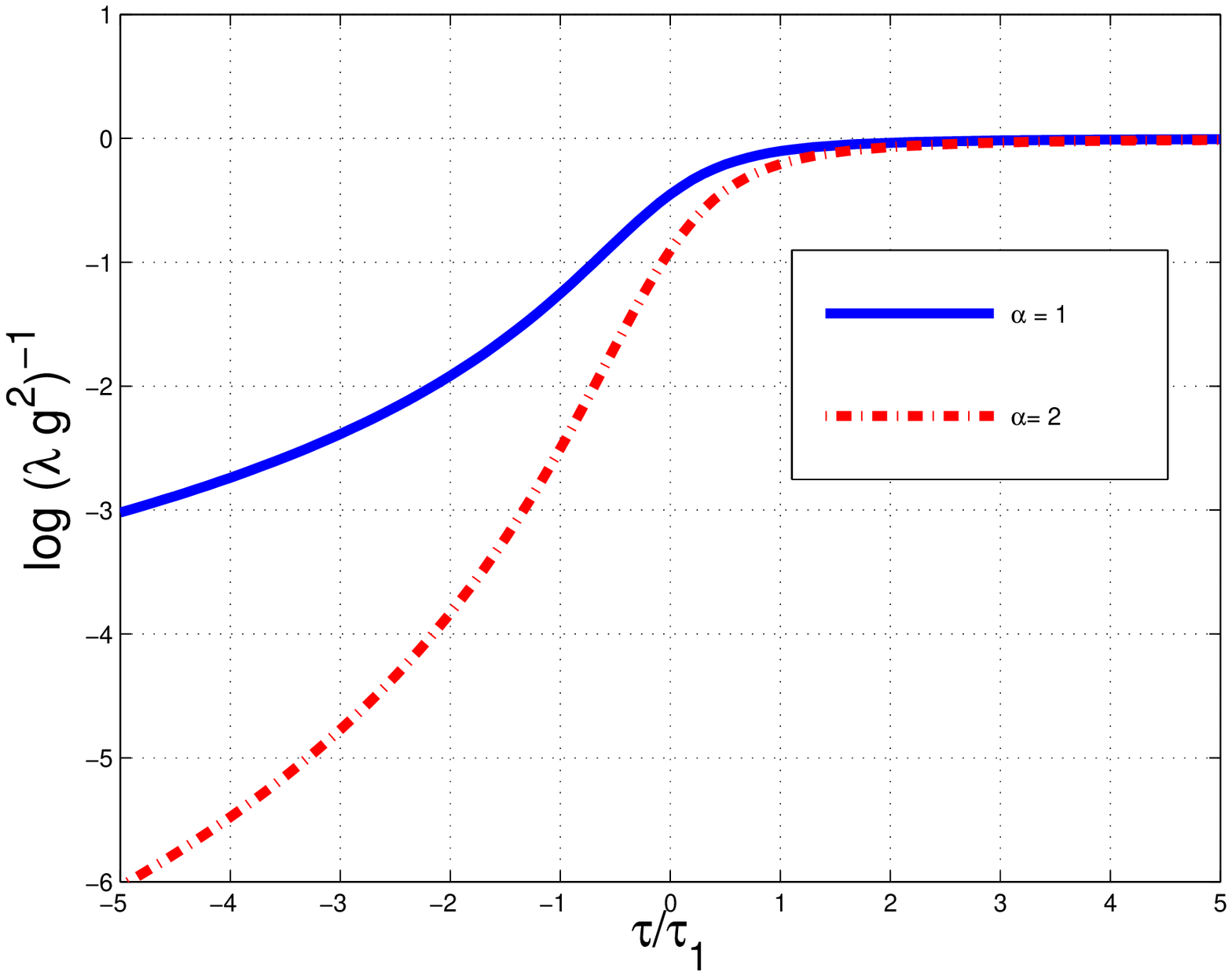}} &
      \hbox{\epsfxsize = 6.9 cm  \epsffile{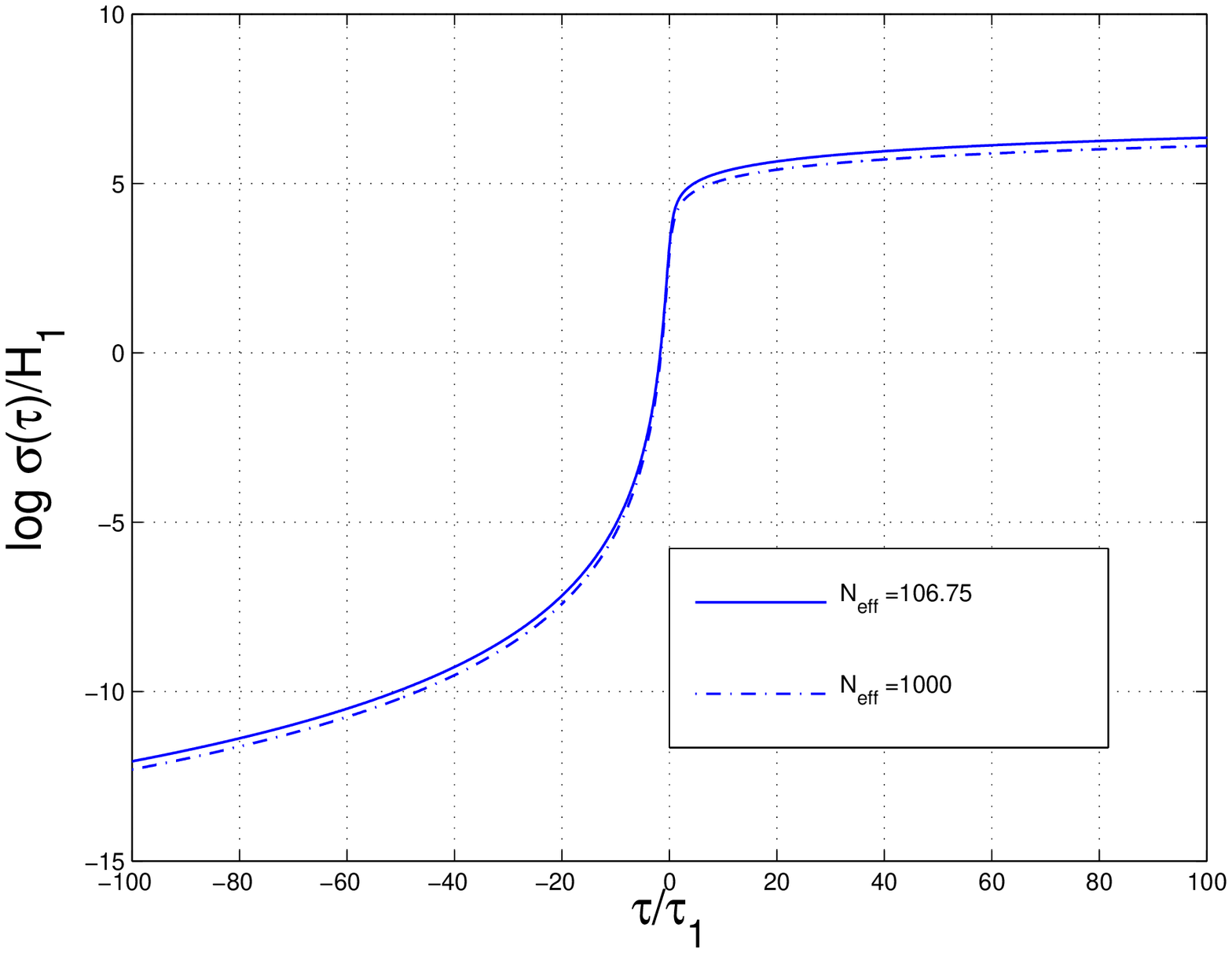}}\\
      \hline
\end{tabular}
\end{center}
\caption{The evolution of $\lambda$ (plot at the left) and of $\sigma$ (plot at the right) is reported 
for different parameters of the model.} 
\label{FIGURE1}
\end{figure}
The analytical calculation will now be corroborated by the appropriate numerical treatment 
where the transition from inflation to radiation is parametrized by
\begin{equation}
a(\tau) = a_{1} ( x + \sqrt{x^2 + 1}),\qquad g^2 \lambda(x) = \biggl(\frac{2 \sqrt{x^2 +1}}{\sqrt{x^2 + 1} + x} \biggr)^{\frac{3\alpha}{2}},\qquad x = \frac{\tau}{\tau_{1}},
\label{M18}
\end{equation}
where $g$ denotes the hypercharge coupling constant. The time $\tau_{1}$ controls 
the duration of the transition regime: for $\tau \ll -\tau_{1}$, $\lambda\simeq (- \tau)^{3 \alpha}$ as implied (to leading order in the slow-roll corrections) by the third relation in Eq. (\ref{Y1})  in conjunction with
 Eq. (\ref{PSI3}). Similarly, if $\tau \ll -\tau_{1}$ the scale factor appearing in Eq. (\ref{M18}) goes as $a(\tau) \simeq (-\tau_{1}/\tau)$ (quasi de-Sitter expansion). Conversely, if   $\tau \gg \tau_{1}$, $a(\tau) \simeq (\tau/\tau_{1})$ (radiation 
 dominated evolution) and $g^2 \lambda\to 1$. 
 The evolution of $\lambda$ is graphically illustrated in Fig. \ref{FIGURE1}) (plot at the left). 
 The time evolution of the conductivity can be modeled as
 \begin{equation}
 \sigma_{\mathrm{c}}(x) = \frac{T_{\mathrm{rh}}}{\alpha} \theta(x),\qquad
 \theta(x) = \frac{1}{8} \biggl( 1 + \frac{x}{\sqrt{x^2 + 1}}\biggr)^{3},
 \label{SM5}
 \end{equation}
where $\theta(x)$ is a smooth representation of the Heaviside step function. Notice 
that the rationale for the third power stems from the fact that 
$\sigma_{\mathrm{c}}(\tau)$ should vanish fast enough for $\tau\ll -\tau_{1}$. The graphic illustration of the evolution 
of $\sigma$ is reported in  Fig. \ref{FIGURE1} (plot at the right).
When the electroweak symmetry is unbroken the conductivity $\sigma_{\mathrm{c}}$
is of the order of $ T/\alpha$ where $\alpha= g^2/4\pi$ and $T$ is the temperature at the corresponding 
epoch. More accurate 
estimates of this quantity exist (see, for instance, \cite{max4} and 
\cite{enqv}) and they agree, up to numerical factors, with the 
figures used in the present paper. In fact, $\sigma_{\mathrm{c}}$ is anyway much larger than the Hubble rate at the corresponding epoch. By relying on the assumption that all the inflaton energy density is efficiently converted into radiation energy density and by assuming a generic number 
$N_{\mathrm{eff}}$ of relativistic degrees of freedom $T_{\mathrm{rh}}$ can be estimated as
\begin{equation}
\frac{T_{\mathrm{rh}}}{H_{1}} = \biggl(\frac{45}{4\pi^4 N_{\mathrm{eff}}}\biggr)^{1/4} (\epsilon\,{\mathcal P}_{{\mathcal R}})^{-1/4},\qquad 
{\mathcal P}_{\mathcal R} = \frac{8}{3} \frac{V}{\epsilon M_{\mathrm{P}}^{4}}
\equiv \frac{1}{24 \pi^2}\frac{V}{\epsilon\overline{M}_{\mathrm{P}}^4},
\label{TRH}
\end{equation}
where the slow-roll equation $  3 H_{1}^2 \overline{M}_{\mathrm{P}}^2 \simeq V$  has been used. Even if $N_{\mathrm{eff}} =106.75$ in the standard model, a drastic variation of one order 
of magnitude does not affect crucially $\sigma$ (see also Fig. \ref{FIGURE1}).
\begin{figure}
\begin{center}
\begin{tabular}{|c|c|}
      \hline
      \hbox{\epsfxsize = 7 cm  \epsffile{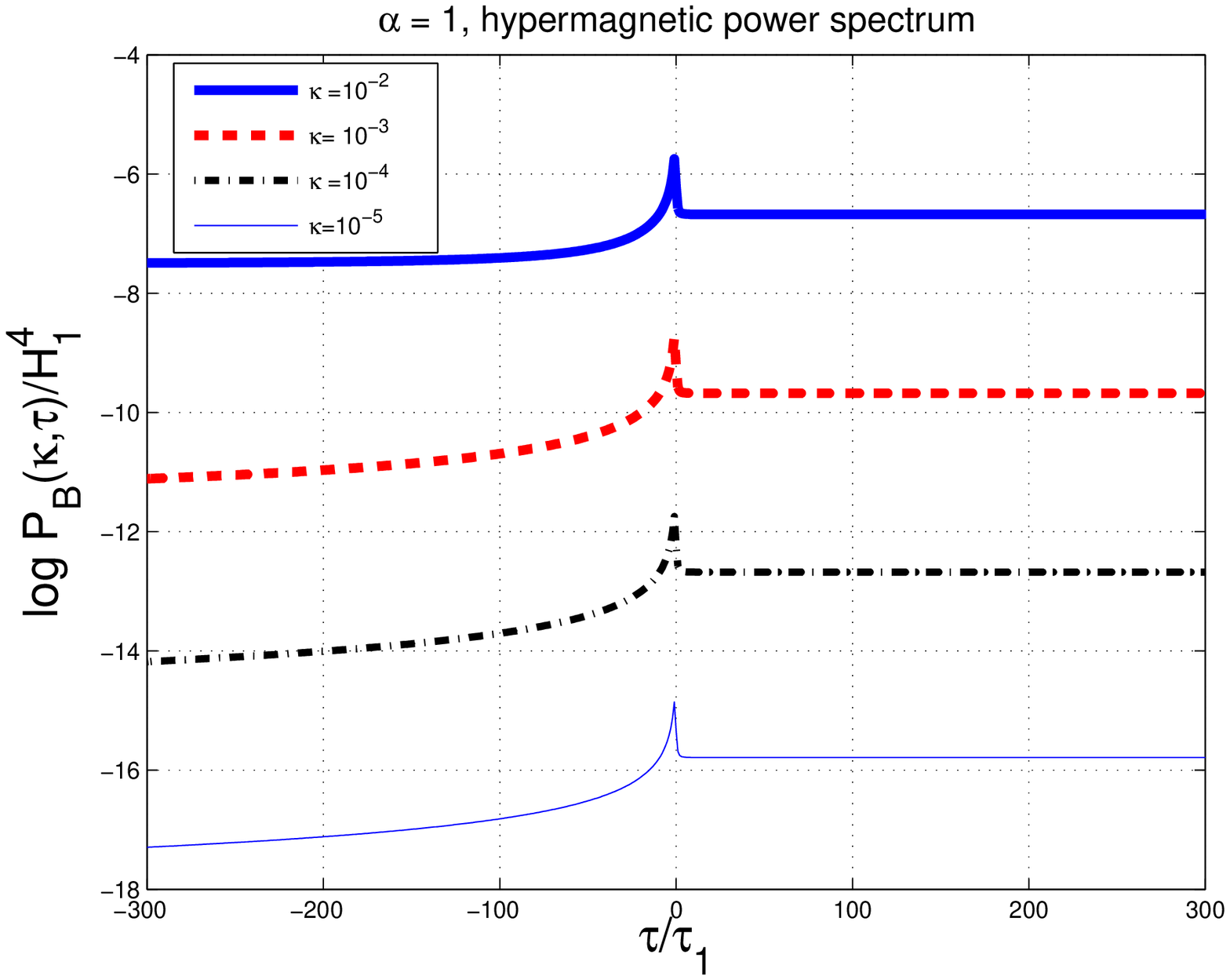}} &
      \hbox{\epsfxsize = 7 cm  \epsffile{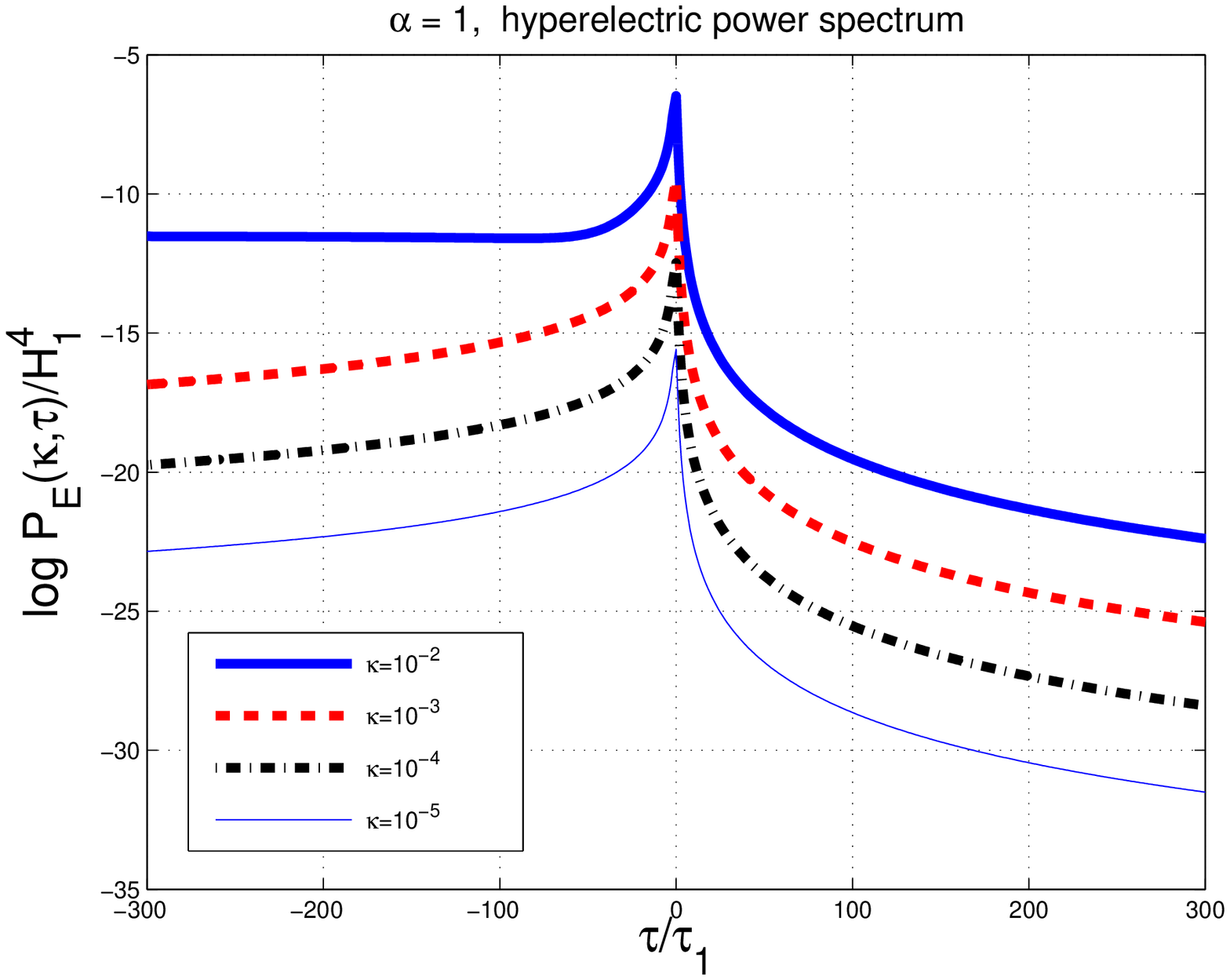}}\\
      \hline
\end{tabular}
\end{center}
\caption{The numerical result for the evolution of the 
hypermagnetic and hyperelectric power spectra
is illustrated on a semi-logarithmic scale.} 
\label{FIGURE2}
\end{figure}
Recalling that $\sigma(\tau)= \sigma_{\mathrm{c}}(\tau) a(\tau)$ the evolution of the mode function in the presence of the Ohmic 
terms 
 \begin{equation}
 f_{k}'' + \frac{4\pi\sigma}{\lambda} f_{k}'+ \biggl\{k^2 - \biggl[ \frac{(\sqrt{\lambda})''}{\sqrt{\lambda}} + 
 \frac{4\pi \sigma}{\lambda} \frac{(\sqrt{\lambda})'}{\sqrt{\lambda}}\biggr]\biggr\} f_{k}=0,
 \end{equation}
 can be solved in the smooth background provided 
 by Eqs. (\ref{M18}) and (\ref{SM5}).
Imposing 
quantum mechanical initial conditions on $f_{k}$  (i.e. $f_{k} = e^{- i k \tau }/\sqrt{2 k}$ for $\tau\to -\infty$) 
the hypermagnetic and hyperelectric power spectra can be obtained and the results are summarized in Fig. \ref{FIGURE2} 
and in the left plot of Fig. \ref{FIGURE3}.
According to Eqs. (\ref{M16}) and (\ref{M16a}), if 
 $\alpha= 1$ (up to slow-roll corrections) $n_{\mathrm{B}} \simeq 4$. Similarly, if $\alpha=2$, $n_{\mathrm{B}}\simeq 1$ 
and the magnetic power spectrum is nearly scale-invariant.
In Fig. \ref{FIGURE2} the hyperelectric and hypermagnetic 
power spectra have been numerically computed in the case 
$\alpha = 1$ and for different values of $\kappa = k\tau_{1}$, i.e. the comoving wave-number in units 
of the transition time $\tau_{1}$. 
The initial integration time $x_{i} =\tau_{i}/\tau_{1}$ depends on the mode and it is chosen in such a way that $\kappa x_{i}> 1$ at $x_{i}$ so that each mode starts its evolution 
inside the Hubble radius. 
By comparing the corresponding values of $\kappa$ in the left and right plots
of Fig. \ref{FIGURE2} the hypermagnetic power spectrum is amplified while the hyperelectric power spectrum is exponentially suppressed.  
\begin{figure}
\begin{center}
\begin{tabular}{|c|c|}
      \hline
      \hbox{\epsfxsize = 7 cm  \epsffile{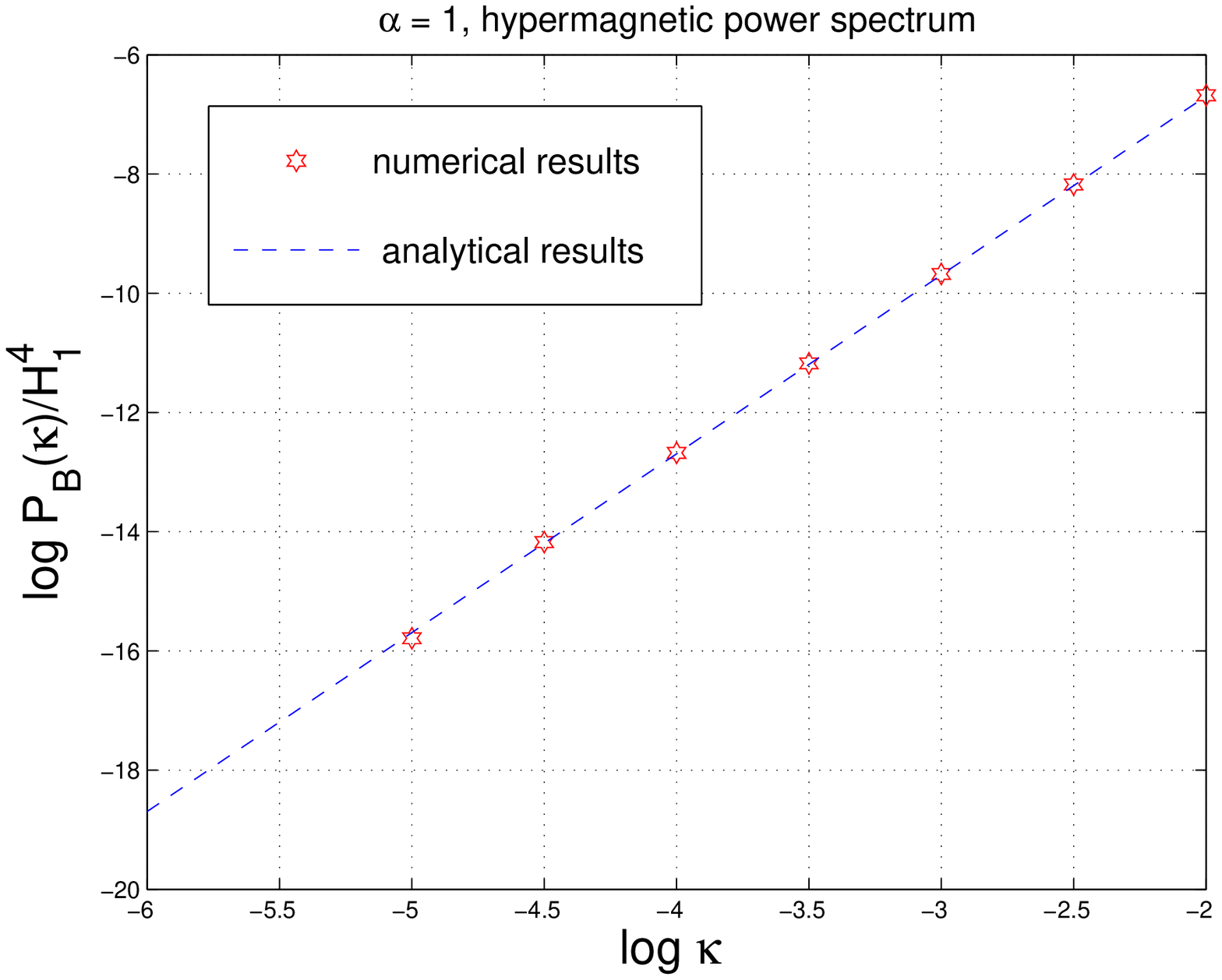}} &
      \hbox{\epsfxsize = 7 cm  \epsffile{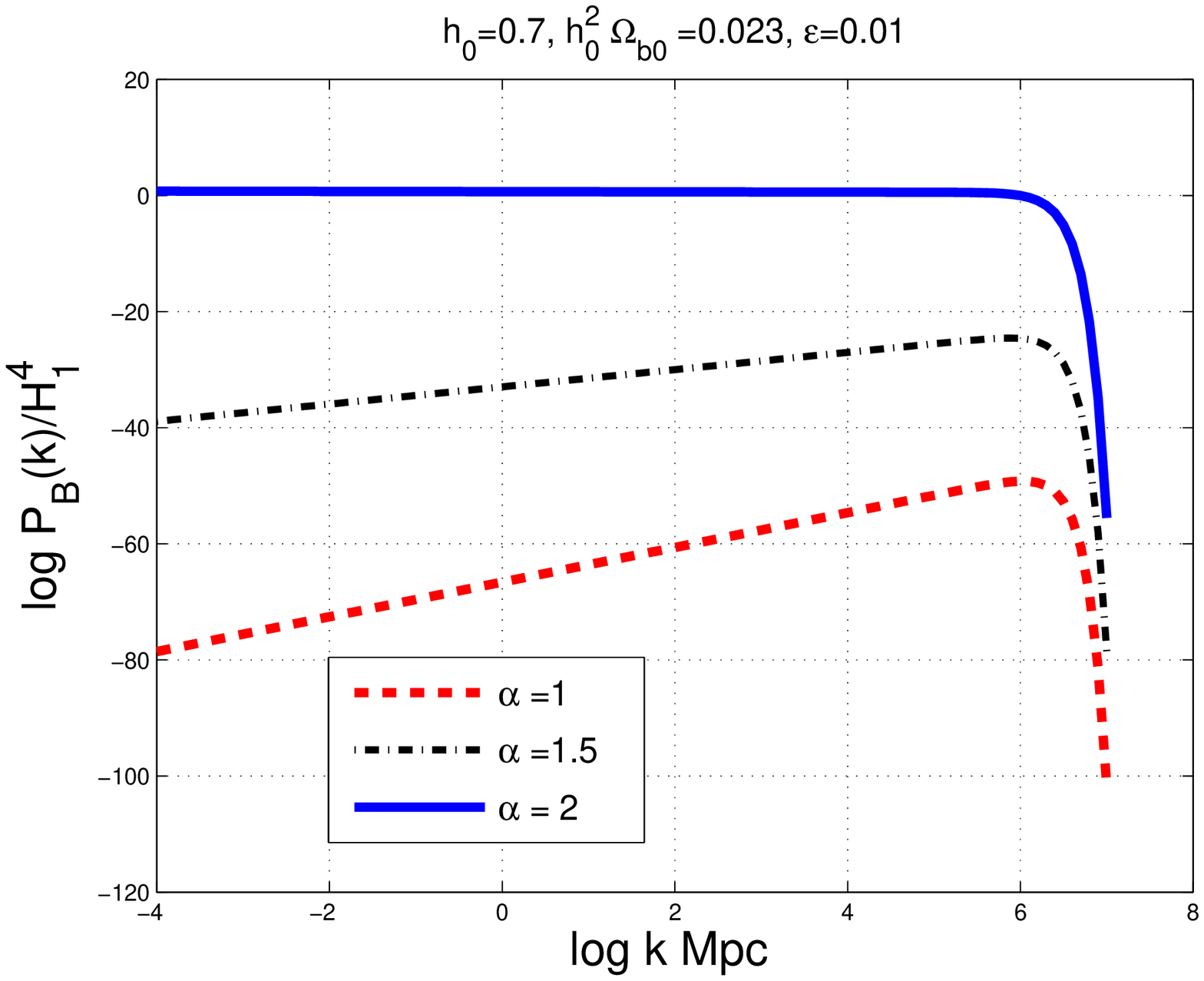}}\\
      \hline
\end{tabular}
\end{center}
\caption{Comparison between analytical and numerical 
results in the case $\alpha=1$ and $\epsilon =0.01$ (plot at the left). The hypermagnetic spectrum  as a function of the comoving 
wave-number in units of $\mathrm{Mpc}^{-1}$ (plot at the right).} 
\label{FIGURE3}
\end{figure}
In Fig. \ref{FIGURE2} (plot at the left) the magnetic power spectrum is reported for different values of the 
wave-number. The amplitude increases with $\kappa$,
which is exactly what we expect in the case of $\alpha=1$ 
where the magnetic power spectrum 
should scale, approximately, as $\kappa^{n_{\mathrm{B}} -1}$ with $n_{\mathrm{B}} = 4$. 
In Fig. \ref{FIGURE3} (plot at the left) a more detailed comparison is illustrated. The starred points correspond to results 
of the numerical integration for different values of the $\kappa$ while the dashed line 
corresponds to the analytical result. From Eqs. (\ref{M16}) and (\ref{M16a}), in the case $\delta = 1$, we obtain
\begin{equation}
\log{\biggl[\frac{P_{\mathrm{B}}(\kappa)}{H_{1}^4}\biggr]}= \log{2} - 2 \log{\pi} + \frac{3  - 4\epsilon}{1-\epsilon} \log{\kappa},
\label{FIT1}
\end{equation}
where we identified 
$k_{1} \simeq \tau_{1}^{-1}$ so that $k/k_{1} \simeq \kappa$.
The dashed line in the left plot of Fig. \ref{FIGURE3} 
is not a fit but it is the result of the analytical expectation. 
Similar agreement is reached for different values of $\alpha$.
Consequently, the analytical results based on the
sudden approximation in conjunction with the matching 
conditions expressed by  Eq. (\ref{M3}) 
are in good agreement with the numerical integration across 
a smooth transition of the same system of equations. 

When the hypermagnetic fields will reenter the Hubble radius (prior to equality but after neutrino decoupling, taking place around the MeV) the electroweak symmetry is already broken. The non-screened vector modes of the 
hypercharge field will the project on the electromagnetic fields as 
${\mathcal A}_{i}^{\mathrm{em}} = \cos{\theta_{\mathrm{w}}} {\mathcal Y}_{i}$. The final magnetic power spectrum can then 
be presented (see Fig. \ref{FIGURE3}, plot at the right) in units of 
$H_{1}^4$, i.e. the fourth power of the Hubble rate at the end of inflation.  
\begin{figure}
\begin{center}
\begin{tabular}{|c|c|}
      \hline
      \hbox{\epsfxsize = 7 cm  \epsffile{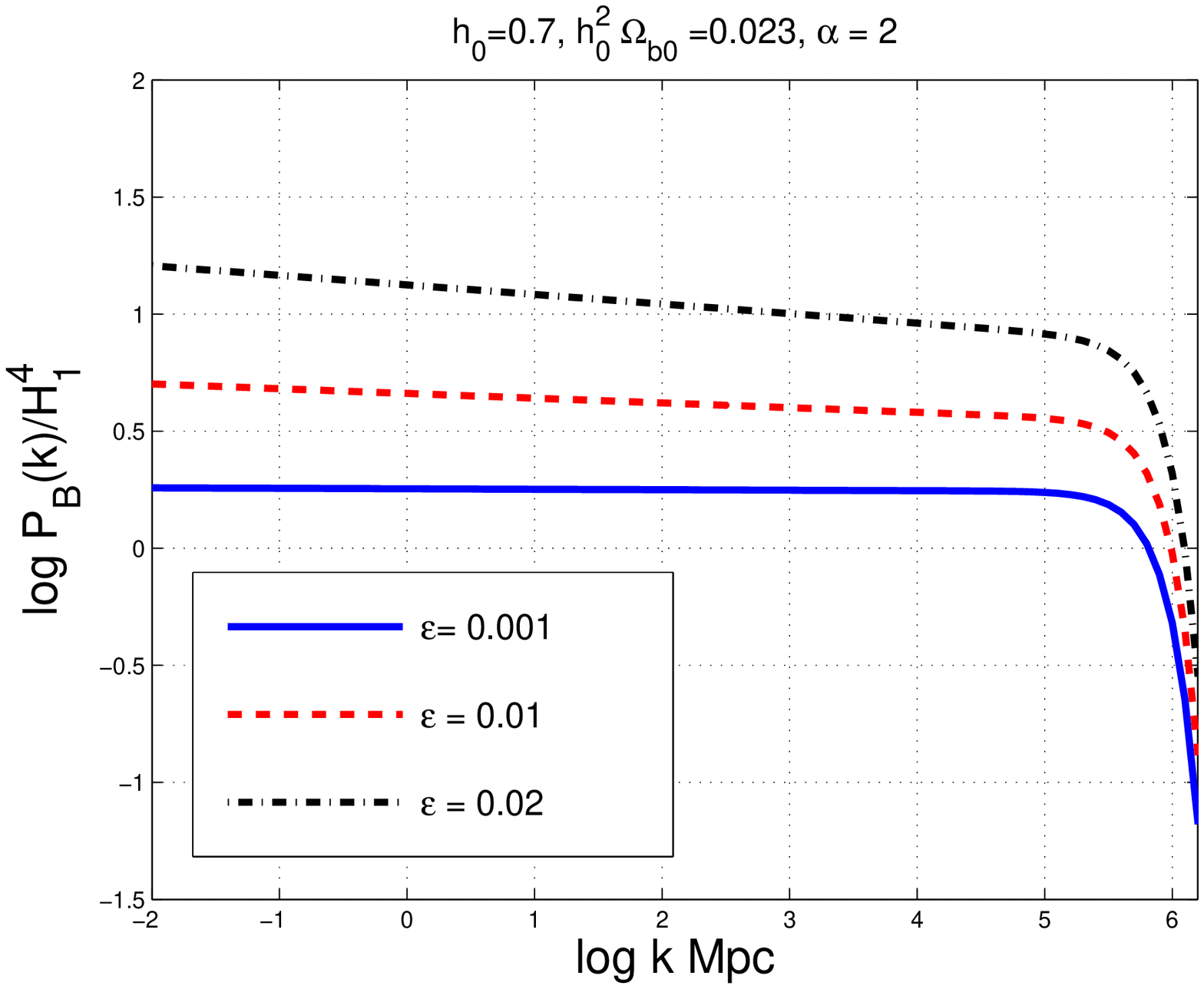}} &
      \hbox{\epsfxsize = 7 cm  \epsffile{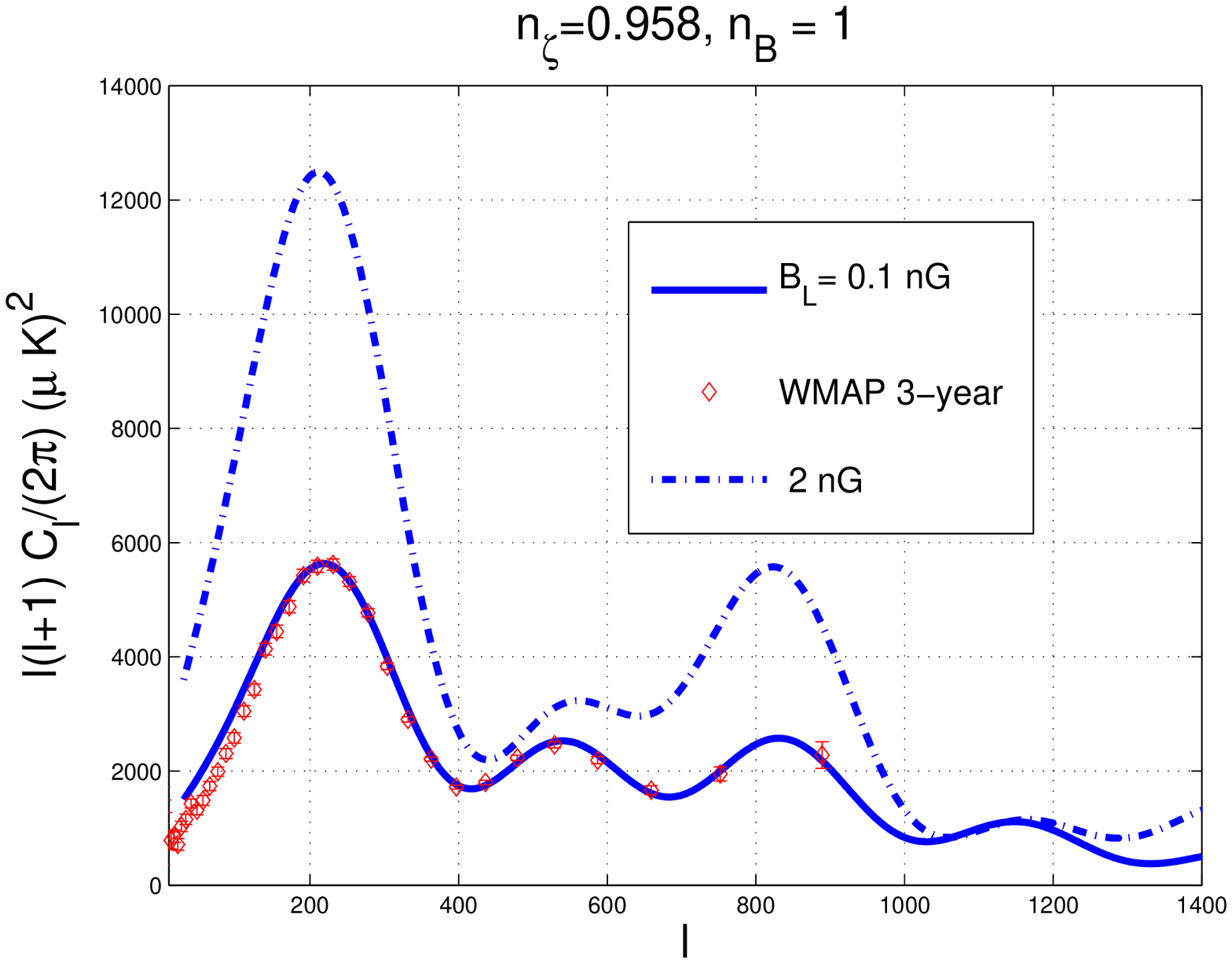}}\\
      \hline
\end{tabular}
\end{center}
\caption{The hypermagnetic power spectrum for different choices of the  parameters and as a function of the comoving wave-number in units of $\mathrm{Mpc}^{-1}$ (plot at the left). The temperature autocorrelations 
of the CMB anisotropies computed in the nearly scale-invariant limit according to the technique described 
in \cite{max3} (see, in particular, third reference).} 
\label{FIGURE4}
\end{figure}
A more physical  measure of the value 
of the obtained magnetic fields is the radiation energy density. The
magnetic power spectrum in units of the radiation background is then
\begin{equation}
\frac{P_{\mathrm{B}}(k)}{8\pi \rho_{\gamma}} = \pi \cos^2{\theta_{\mathrm{w}}} {\cal C}(\delta) \,\epsilon\,{\mathcal P}_{\mathcal R} \biggl(
\frac{k}{k_{1}}\biggr)^{n_{\mathrm{B}} -1},
\label{magnrad}
\end{equation}
where both $n_{\mathrm{B}}$ and $\delta$ depend upon the slow-roll parameter $\epsilon$. Since \cite{WMAP} ${\mathcal P}_{{\mathcal R}} \simeq 2.35 \times 10^{-9}$, in the scale-invariant limit Eq. (\ref{magnrad}) is of the order of $10^{-10}$. 
Consequently, the
present value of the magnetic field is of order $0.1$--$0.01$ nG with a theoretical error 
that depends upon $\epsilon$ (which should be smaller than about $0.05$ according to 
current experimental data\footnote{It is possible to obtain an upper bound on $\epsilon$ by 
analyzing, for instance, CMB and large-scale structure data within a $\Lambda$CDM model 
containing also a tensor component. The analysis will then lead to an upper limit on the 
ratio between tensor and scalar power spectra which can be translated into an upper limit on $\epsilon$. The combination of WMAP \cite{WMAP} data and the data of the Sloan Digital Sky Survey \cite{SDSS} would lead, for instance to $\epsilon \laq  0.02$}). 

The results of Eq. (\ref{magnrad}) can also be illustrated by regularizing 
the magnetic field over a typical comoving scale $L$ by means of a Gaussian window function in Fourier space
\cite{max3,tin}. Denoting as $B_{\mathrm{L}}$ the regularized magnetic field over the comoving scale 
$L = 2\pi/k_{\mathrm{L}}$ we will have, in the nearly scale-invariant limit and at the time of the collapse of the protogalaxy, 
\begin{equation}
\biggl(\frac{B_{\mathrm{L}}}{\mathrm{nG}}\biggr) \simeq 0.1 \biggl(\frac{\epsilon}{0.01}\biggr)^{1/2} \biggl(\frac{{\mathcal P}_{\mathcal R}}{2.35 \times 10^{-9}}\biggr)^{1/2}.
\label{magnreg}
\end{equation}
The magnetic energy density in units 
of the radiation background can be expressed, in this case, as 
\begin{equation}
\overline{\Omega}_{\mathrm{BL}} = \frac{B_{\mathrm{L}}^2}{8\pi \rho_{\gamma}} = 
7.56 \times 10^{-9} \, \biggl(\frac{B_{\mathrm{L}}}{\mathrm{nG}}\biggr)^2,
\label{OM1}
\end{equation}
where the pivot value of $B_{\mathrm{L}}$ has been taken at the epoch of gravitational collapse 
of the protogalaxy.
It is customary 
to require, for a successful magnetogenesis \cite{weyl1}, that\footnote{The first of these two figures stems from 
the (overoptimistic) requirement that the galactic 
dynamo is so efficient to amplify the protogalactic field by one e-fold for each galactic rotation. Strictly speaking this argument only applies to spiral galaxies. The second requirement 
takes into account the possible early saturation of galactic dynamo and it is still rather optimistic (see, for instance, the second reference of \cite{zel1} for a discussion). } $\overline{\Omega}_{\mathrm{BL}} > 10^{-34}$, or, more realistically $\overline{\Omega}_{\mathrm{BL}} > 10^{-24}$. 

Equation (\ref{magnreg}) is compatible with a galactic magnetic field of the order of the $\mu$G.
During the process 
of collapse the magnetic flux is frozen into the plasma element thanks to the large value of the conductivity. The mean matter 
density increases, during collapse,
from its critical value (i.e. $\rho_{\mathrm{cr}} =1.05\times 10^{-5} \,h_{0}^2 \mathrm{GeV}/\mathrm{cm}^3$) to a final value $\rho_{\mathrm{f}}$ value which is 5 to 6 orders of magnitude larger than $\rho_{\mathrm{c}}$. The magnetic field after collapse will then be
\begin{equation}
B_{\mathrm{gal}} = \biggl(\frac{\rho_{\mathrm{f}}}{\rho_{\mathrm{c}}}\biggr)^{2/3} B_{\mathrm{L}} \simeq \biggl(\frac{\epsilon}{0.01}\biggr)^{1/2} \biggl(\frac{{\mathcal P}_{\mathcal R}}{2.35 \times 10^{-9}}\biggr)^{1/2}\, \mu\mathrm{G}.
\end{equation}
Over present length-scales much larger than the Mpc the magnetic fields, in this model, will be (today) smaller than the nG since these regions did not benefit of the compressional amplification. Within this lore the magnetic fields in clusters could be produced by magnetic reconnection from the ones of the galaxies but the experimental uncertainty in their correlation scale \cite{clust1} does not allow a definite statement. If 
the spectrum of the primeval field is not nearly scale-invariant  its amplitude over a comoving Mpc scale will be smaller (see right plot of Fig. \ref{FIGURE3}) and, consequently, a non negligible 
dynamo action will be required for the phenomenological relevance of the obtained result. In this second scenario the 
cluster magnetic field might be related to the way the dynamo is saturated.

In a series of papers a semi-analytical technique has been developed 
for the evaluation of the temperature autocorrelations 
(see, in particular, the third reference in \cite{max3}). Since 
in the present model the cross-correlation between 
magnetic and adiabatic contribution vanishes the temperature 
cross-correlations are given, for multipoles $\ell < 30$ by 
the following generalization of the Sachs-Wolfe plateau:
 \begin{eqnarray}
C^{(\mathrm{SW})}_{\ell} &=& \biggl[ \frac{{\cal P}_{\mathcal R}}{25} \,{\mathcal Z}_{1}(n,\ell)  +
\frac{\epsilon^2 {\mathcal P}^2_{\mathcal R}}{400} \, R_{\gamma}^2   {\mathcal Z}_{2}(n_{\mathrm{B}},\ell) \biggr],
\label{SW1}\\
{\mathcal Z}_{1}(n,\ell) &=& \frac{\pi^2}{4} \biggl(\frac{k_0}{k_{\rm p}}\biggl)^{n-1} 2^{n} \frac{\Gamma( 3 - n) \Gamma\biggl(\ell + 
\frac{ n -1}{2}\biggr)}{\Gamma^2\biggl( 2 - \frac{n}{2}\biggr) \Gamma\biggl( \ell + \frac{5}{2} - \frac{n}{2}\biggr)},
\label{Z1}\\
{\mathcal Z}_{2}(n_{\mathrm{B}},\ell) &=& \frac{\pi^2}{2} 2^{2(n_{\mathrm{B}} -1)} {\cal F}(n_{\mathrm{B}}) \biggl( \frac{k_{0}}{k_{1}}\biggr)^{ 2(n_{\mathrm{B}} -1)} \frac{ \Gamma( 4 - 2 n_{\mathrm{B}}) 
\Gamma(\ell + n_{\mathrm{B}} -1)}{\Gamma^2\biggl(\frac{5}{2} - n_{\mathrm{B}}\biggr) 
\Gamma(\ell + 3 - n_{\mathrm{B}})},
\label{Z2}\\
{\mathcal   F}(n_{\mathrm{B}}) &=& \frac{4 \pi^2}{27} {\mathcal C}^2(\delta)\frac{(7 - n_{\mathrm{B}}) }{(n_{\mathrm{B}} -1) ( 5 - 2 n_{\mathrm{B}})
},\qquad n_{\mathrm{B}}>1,
 \label{Fnb}
\end{eqnarray}
where $n$ denotes the spectral index of the adiabatic mode\footnote{For the numerical estimate of Fig. \ref{FIGURE4}, $n$ will be taken to be $0.958$ which is the best fit value obtainable by analyzing the 
WMAP data alone \cite{WMAP}.}. If $n_{\mathrm{B}}\simeq 1$ 
the function (\ref{Fnb}) contains the logarithm of the infra-red cut-off.

At smaller angular scales (i.e. $\ell > 100$) the temperature 
autocorrelations can be obtained, within the scheme of \cite{max3} 
(third reference) by computing numerically four integrals
 and the results for multipoles compatible with the Doppler oscillations 
are reported in Fig. \ref{FIGURE4} (plot at the right).  When $B_{\mathrm{L}} \simeq 0.1 $ nG the structure of the Doppler oscillations is not altered (see also \cite{max3} for a model-independent discussion). 

So far it has been assumed that the decay rate of the inflaton
and of the spectator fields were comparable, i.e. $\Gamma_{\psi} \simeq \Gamma_{\varphi}$. It can also happen that 
the spectator field decays later than the inflaton. The predicted slopes of the magnetic power spectra will not be modified for comoving scales of the order of the Mpc. However, shorter wavelengths can be affected if the decay of $\psi$ is delayed.
In the latter case since $\psi_{*} < \overline{M}_{\mathrm{P}}$, the ratio between 
the energy density of $\psi$ and the radiation background (produced by the inflaton decay) will 
grow, after inflation.  The fluctuations of the spectator field 
may then represent a further source of curvature perturbations.
If the inflationary Hubble rate is much smaller than $10^{-5} \overline{M}_{\mathrm{P}}$ the fluctuations of $\psi$ will eventually become the dominant source of curvature perturbations as noticed in the context of the 
so-called curvaton scenario \cite{curv1}. In the opposite case the two contributions will interfere.
In the latter case the final spectrum 
of curvature perturbations can be computed, for different post-inflationary 
evolutions, as a function of $\psi_{*}$. This analysis can be carried on numerically with the techniques already exploited in 
\cite{curv2}. The 
final result can be written in terms of the amplitude 
of the curvature perturbations at the pivot scale:
\begin{equation}
{\mathcal P}_{\mathcal R} = \frac{1}{24\pi^2} \frac{V}{\overline{M}_{\mathrm{P}}^4}\biggl[ \frac{1}{\epsilon} + f^2(\psi_{*})\biggr],\qquad 
f(\psi_{*}) = c_{1} \biggl(\frac{\psi_{*}}{\overline{M}_{\mathrm{P}}}\biggr) 
+ c_{2} \biggl(\frac{\overline{M}_{\mathrm{P}}}{\psi_{*}}\biggr)
\label{CC1}
\end{equation}
where $c_{1}= 0.13$ and $c_2= 0.25$. In the limit $f\to 0$ 
we recover the result of Eq. (\ref{TRH}). In the case $f\neq 0$ 
the curvature fluctuations induced by $\psi$ may mix, in the Sachs-Wolfe 
plateau, with the component induced by the inflaton fluctuations. 
In some cases there could even be a correlation term. As argued 
in \cite{curv2} (second reference) these results strongly depend upon 
$W(\psi)$ being quadratic and not, for instance quartic. In spite 
of the details of the post-inflationary history Eq. (\ref{CC1}) suggests a possible violation of the consistency relation 
which would become, in the case of Eq. (\ref{CC1}), 
$r_{\mathrm{T}} = 16 \epsilon/[1 + 8 f^2(\psi_{*}) \epsilon]$ having 
defined  $r_{\mathrm{T}} = {\mathcal P}_{\mathrm{T}}/{\mathcal P}_{\mathcal R}$, i.e. the ratio between the tensor and the scalar 
power spectra. This prediction would allow, in principle, to distinguish observationally 
the situations where the spectator field decays during reheating or later. We leave for a 
forthcoming paper the detailed analysis of this and other cases \cite{mg}. 

The main goal of the present study  has been 
 to demonstrate, within conventional inflationary scenarios,  the  viability  of a class of magnetogenesis models 
 that do not require a strong dynamo action and that 
are compatible, at the same time,  with the direct bounds stemming from the analysis of the CMB anisotropies.  
The foreseeable improvement  of the quality of CMB data stimulates 
the effort of more accurate calculations of the impact of a magnetized plasma 
upon the various CMB observables. As explicitly demonstrated in this paper 
is possible to construct viable magnetogenesis models which have well defined CMB signatures.
Since theoretical prejudices (and diatribes)  
are not a decisive proof for the existence (or not existence) of pre-recombination magnetic fields, it is 
wise pursue the development of model-independent tools for accurate analyses of magnetized CMB anisotropies,
as suggested by the present investigation. 
Indeed, forthcoming satellite experiments may turn some of the present speculations in more solid scientific 
statements either in favour or against the primordial hypothesis.

\end{document}